\input eplain
\input amssym.def


\magnification=\magstephalf
\hsize=14.0 true cm
\vsize=19 true cm
\hoffset=1.0 true cm
\voffset=2.0 true cm

\abovedisplayskip=12pt plus 3pt minus 3pt
\belowdisplayskip=12pt plus 3pt minus 3pt
\parindent=1.0em


\leftdisplays
\leftdisplayindent=1.0em


\font\sixrm=cmr6
\font\eightrm=cmr8
\font\ninerm=cmr9

\font\sixi=cmmi6
\font\eighti=cmmi8
\font\ninei=cmmi9

\font\sixsy=cmsy6
\font\eightsy=cmsy8
\font\ninesy=cmsy9

\font\sixbf=cmbx6
\font\eightbf=cmbx8
\font\ninebf=cmbx9

\font\eightit=cmti8
\font\nineit=cmti9

\font\eightsl=cmsl8
\font\ninesl=cmsl9

\font\sixss=cmss8 at 8 true pt
\font\sevenss=cmss9 at 9 true pt
\font\eightss=cmss8
\font\niness=cmss9
\font\tenss=cmss10

 at 12 true pt
\font\bigrm=cmr10 at 12 true pt
 at 12 true pt

 at 14 true pt
\font\Bigrm=cmr12 at 16 true pt
 at 14 true pt

\catcode`@=11
\newfam\ssfam

\def\tenpoint{\def\rm{\fam0\tenrm}%
    \textfont0=\tenrm \scriptfont0=\sevenrm \scriptscriptfont0=\fiverm
    \textfont1=\teni  \scriptfont1=\seveni  \scriptscriptfont1=\fivei
    \textfont2=\tensy \scriptfont2=\sevensy \scriptscriptfont2=\fivesy
    \textfont3=\tenex \scriptfont3=\tenex   \scriptscriptfont3=\tenex
    \textfont\itfam=\tenit                  \def\it{\fam\itfam\tenit}%
    \textfont\slfam=\tensl                  \def\sl{\fam\slfam\tensl}%
    \textfont\bffam=\tenbf \scriptfont\bffam=\sevenbf
    \scriptscriptfont\bffam=\fivebf
                                            \def\bf{\fam\bffam\tenbf}%
    \textfont\ssfam=\tenss \scriptfont\ssfam=\sevenss
    \scriptscriptfont\ssfam=\sevenss
                                            \def\ss{\fam\ssfam\tenss}%
    \normalbaselineskip=13pt
    \setbox\strutbox=\hbox{\vrule height8.5pt depth3.5pt width0pt}%
    \let\big=\tenbig
    \normalbaselines\rm}

\def\ninepoint{\def\rm{\fam0\ninerm}%
    \textfont0=\ninerm      \scriptfont0=\sixrm
                            \scriptscriptfont0=\fiverm
    \textfont1=\ninei       \scriptfont1=\sixi
                            \scriptscriptfont1=\fivei
    \textfont2=\ninesy      \scriptfont2=\sixsy
                            \scriptscriptfont2=\fivesy
    \textfont3=\tenex       \scriptfont3=\tenex
                            \scriptscriptfont3=\tenex
    \textfont\itfam=\nineit \def\it{\fam\itfam\nineit}%
    \textfont\slfam=\ninesl \def\sl{\fam\slfam\ninesl}%
    \textfont\bffam=\ninebf \scriptfont\bffam=\sixbf
                            \scriptscriptfont\bffam=\fivebf
                            \def\bf{\fam\bffam\ninebf}%
    \textfont\ssfam=\niness \scriptfont\ssfam=\sixss
                            \scriptscriptfont\ssfam=\sixss
                            \def\ss{\fam\ssfam\niness}%
    \normalbaselineskip=12pt
    \setbox\strutbox=\hbox{\vrule height8.0pt depth3.0pt width0pt}%
    \let\big=\ninebig
    \normalbaselines\rm}

\def\eightpoint{\def\rm{\fam0\eightrm}%
    \textfont0=\eightrm      \scriptfont0=\sixrm
                             \scriptscriptfont0=\fiverm
    \textfont1=\eighti       \scriptfont1=\sixi
                             \scriptscriptfont1=\fivei
    \textfont2=\eightsy      \scriptfont2=\sixsy
                             \scriptscriptfont2=\fivesy
    \textfont3=\tenex        \scriptfont3=\tenex
                             \scriptscriptfont3=\tenex
    \textfont\itfam=\eightit \def\it{\fam\itfam\eightit}%
    \textfont\slfam=\eightsl \def\sl{\fam\slfam\eightsl}%
    \textfont\bffam=\eightbf \scriptfont\bffam=\sixbf
                             \scriptscriptfont\bffam=\fivebf
                             \def\bf{\fam\bffam\eightbf}%
    \textfont\ssfam=\eightss \scriptfont\ssfam=\sixss
                             \scriptscriptfont\ssfam=\sixss
                             \def\ss{\fam\ssfam\eightss}%
    \normalbaselineskip=10pt
    \setbox\strutbox=\hbox{\vrule height7.0pt depth2.0pt width0pt}%
    \let\big=\eightbig
    \normalbaselines\rm}

\def\tenbig#1{{\hbox{$\left#1\vbox to8.5pt{}\right.\n@space$}}}
\def\ninebig#1{{\hbox{$\textfont0=\tenrm\textfont2=\tensy
                       \left#1\vbox to7.25pt{}\right.\n@space$}}}
\def\eightbig#1{{\hbox{$\textfont0=\ninerm\textfont2=\ninesy
                       \left#1\vbox to6.5pt{}\right.\n@space$}}}

\font\sectionfont=cmbx10
\font\subsectionfont=cmti10

\def\figurecaptionfont{\ninepoint}
\def\tablecaptionfont{\ninepoint}
\def\footnotefont{\eightpoint}


\newcount\equationno
\newcount\bibitemno
\newcount\figureno
\newcount\tableno

\equationno=0
\bibitemno=0
\figureno=0
\tableno=0


\footline={\ifnum\pageno=0{\hfil}\else
{\hss\rm\the\pageno\hss}\fi}


\def\section #1. #2 \par
{\vskip0pt plus .20\vsize\penalty-100 \vskip0pt plus-.20\vsize
\vskip 1.6 true cm plus 0.2 true cm minus 0.2 true cm
\global\def\equationlabel{#1}
\global\equationno=0
\leftline{\sectionfont #1. #2}\par
\immediate\write\terminal{Section #1. #2}
\vskip 0.7 true cm plus 0.1 true cm minus 0.1 true cm
\noindent}


\def\subsection #1 \par
{\vskip0pt plus 0.8 true cm\penalty-50 \vskip0pt plus-0.8 true cm
\vskip2.5ex plus 0.1ex minus 0.1ex
\leftline{\subsectionfont #1}\par
\immediate\write\terminal{Subsection #1}
\vskip1.0ex plus 0.1ex minus 0.1ex
\noindent}


\def\appendix #1 \par
{\vskip0pt plus .20\vsize\penalty-100 \vskip0pt plus-.20\vsize
\vskip 1.6 true cm plus 0.2 true cm minus 0.2 true cm
\global\def\equationlabel{\hbox{\rm#1}}
\global\equationno=0
\leftline{\sectionfont Appendix #1}\par
\immediate\write\terminal{Appendix #1}
\vskip 0.7 true cm plus 0.1 true cm minus 0.1 true cm
\noindent}


\def\enum{\global\advance\equationno by 1
(\equationlabel.\the\equationno)}


\def\ifundefined#1{\expandafter\ifx\csname#1\endcsname\relax}

\def\ref#1{\ifundefined{#1}?\immediate\write\terminal{unknown reference
on page \the\pageno}\else\csname#1\endcsname\fi}

\newwrite\terminal
\newwrite\bibitemlist

\def\bibitem#1#2\par{\global\advance\bibitemno by 1
\immediate\write\bibitemlist{\string\def
\expandafter\string\csname#1\endcsname
{\the\bibitemno}}
\item{[\the\bibitemno]}#2\par}

\def\beginbibliography{
\vskip0pt plus .15\vsize\penalty-100 \vskip0pt plus-.15\vsize
\vskip 1.2 true cm plus 0.2 true cm minus 0.2 true cm
\leftline{\sectionfont References}\par
\immediate\write\terminal{References}
\immediate\openout\bibitemlist=biblist
\frenchspacing\parindent=1.8em
\vskip 0.5 true cm plus 0.1 true cm minus 0.1 true cm}

\def\endbibliography{
\immediate\closeout\bibitemlist
\nonfrenchspacing\parindent=1.0em}


\def\figurecaption#1{\global\advance\figureno by 1
\narrower\figurecaptionfont
Fig.~\the\figureno. #1}

\def\tablecaption#1{\global\advance\tableno by 1
\vbox to 0.5 true cm { }
\centerline{\tablecaptionfont%
Table~\the\tableno. #1}
\vskip-0.4 true cm}

\def\thintablerule{\hrule height0.4pt}

\tenpoint

\immediate\openin\bibitemlist=biblist
\ifeof\bibitemlist\immediate\closein\bibitemlist
\else\immediate\closein\bibitemlist
\input biblist \fi


\def\thismonth{\ifcase\month\or
January\or February\or March\or April\or May\or June\or
July\or August\or September\or October\or November\or December\fi}



\def\rmd{{\rm d}}
\def\rmD{{\rm D}}
\def\rme{{\rm e}}
\def\rmO{{\rm O}}


\def\rz{{\Bbb R}}


\def\proof{\noindent{\sl Proof:}\kern0.6em}

\def\frac#1#2{\hbox{$#1\over#2$}}
\def\dual{\mathstrut^*\kern-0.1em}

\def\lvec#1{\setbox0=\hbox{$#1$}
    \setbox1=\hbox{$\scriptstyle\leftarrow$}
    #1\kern-\wd0\smash{
    \raise\ht0\hbox{$\raise1pt\hbox{$\scriptstyle\leftarrow$}$}}
    \kern-\wd1\kern\wd0}
\def\rvec#1{\setbox0=\hbox{$#1$}
    \setbox1=\hbox{$\scriptstyle\rightarrow$}
    #1\kern-\wd0\smash{
    \raise\ht0\hbox{$\raise1pt\hbox{$\scriptstyle\rightarrow$}$}}
    \kern-\wd1\kern\wd0}


\def\nab#1{{\nabla_{#1}}}
\def\nabstar#1{\nabla\kern-0.5pt\smash{\raise 4.5pt\hbox{$\ast$}}
               \kern-4.5pt_{#1}}

\def\drvstar#1{\partial\kern-0.5pt\smash{\raise 4.5pt\hbox{$\ast$}}
               \kern-5.0pt_{#1}}

\def\ldrvstar#1{\lvec{\,\partial}\kern-0.5pt\smash{\raise 4.5pt\hbox{$\ast$}}
               \kern-5.0pt_{#1}}




\def\psibar{\overline{\psi}}
\def\anomaly{{\cal A}}


\def\dirac#1{\gamma_{#1}}
\def\diracstar#1#2{
    \setbox0=\hbox{$\gamma$}\setbox1=\hbox{$\gamma_{#1}$}
    \gamma_{#1}\kern-\wd1\kern\wd0
    \smash{\raise4.5pt\hbox{$\scriptstyle#2$}}}
\def\dirachat{\hat{\gamma}_5}


\def\group{G}

\def\tr{{\rm tr}}
\def\Tr{{\rm Tr}}

\def\d#1{d^{#1}_{\hbox{$\scriptstyle\kern-0.5pt R\scriptfont1=\sixi$}}}


\def\L{{\frak L}}

\def\Wline{{W}}


\def\trans{{\cal Q}}
\def\vartrans{{\cal S}}


\def\ca{c_1}
\def\ci{c_2}


\def\D{D^A}%
\rightline{DESY 99-040}

\vskip 1.0 true cm minus 0.3 true cm
\centerline
{\Bigrm Weyl fermions on the lattice and the}
\vskip 1.5ex
\centerline
{\Bigrm non-abelian gauge anomaly}
\vskip 0.6 true cm
\centerline{\bigrm Martin L\"uscher\kern1pt%
\footnote{$^{\ast}$}{\footnotefont E-mail: luscher@mail.desy.de}}
\vskip1ex
\centerline{\it Deutsches Elektronen-Synchrotron DESY}
\centerline{\it Notkestrasse 85, D-22603 Hamburg, Germany}
%
\vskip 1.0 true cm
\thintablerule
\vskip 2.0ex
\ninepoint
\leftline{\bf Abstract}
\vskip 1.0ex\noindent
Starting from the Ginsparg-Wilson relation, a general construction of
chiral gauge theories on the lattice is described.
Local and global anomalies are easily discussed in this framework
and a closed expression for the effective action can be obtained.
Particular attention is paid to the non-abelian gauge anomaly,
which is shown to be related to 
a local topological field on the lattice
representing the Chern character in 4+2 dimensions.


\vskip 2.0ex
\thintablerule
\tenpoint

\vskip-0.5 true cm

\section 1. Introduction

In abelian chiral gauge theories the gauge anomaly
is proportional to the topological charge density
and its topological significance is hence relatively 
easy to understand. 
As has recently become 
clear [\ref{HasenfratzI}--\ref{LuscherSymmetry}],
the same is true on the lattice
if the lattice Dirac operator $D$
satisfies the Ginsparg-Wilson relation [\ref{GinspargWilson}]
$$
  \dirac{5}D+D\dirac{5}=aD\dirac{5} D.
  \eqno\enum
$$
For any value of the lattice spacing $a$, 
this identity implies an exact symmetry of the fermion action,
which may regarded as a lattice version of the usual
chiral rotations. Moreover the axial anomaly (which coincides with the 
gauge anomaly in the abelian case) arises from the non-invariance
of the fermion integration measure under these transformations 
and can be shown to be a topological field, 
i.e.~the associated charge does not change under local deformations of the 
gauge field.

This has now led to a construction of 
abelian chiral gauge theories on the lattice,
which complies with all the basic requirements including
exact gauge invariance [\ref{LuscherAbelian}]. 
The fermion multiplet has to be anomaly-free for this to work
out, but otherwise there are no restrictions and 
the anomaly cancellation can be proved on the basis 
of the topological nature of the anomaly alone [\ref{LuscherAnomaly}].

If the gauge group is not abelian, the gauge anomaly assumes a more
complicated form and its topological interpretation is not
immediately clear.
An important clue is provided by the 
Stora-Zumino descent equations [\ref{StoraI}--\ref{Zumino}],
which allow one to pass from the
Chern character in 4+2 dimensions (an expression proportional to the 
third power of the gauge field tensor)
via the Chern-Simons term in 4+1 dimensions to the anomaly in 
4 dimensions. It is then possible to 
show [\ref{AlvarezGinsparg}] that the anomaly is related to
the index theorem in 4+2 dimensions and to  
the existence of certain non-contractible two-spheres in the space of 
gauge orbits
(for a review and an extensive list of references see
refs.~[\ref{EriceLectures},\ref{Bertlmann}])

In this paper a general formulation of chiral gauge theories 
on the lattice is proposed. The basic ansatz is the same
as in the case of the abelian theories considered in 
ref.~[\ref{LuscherAbelian}], but there are some
new elements which make the approach more transparent.
In particular, a direct 
connection between the gauge anomaly and a local topological field
representing the Chern character in 4+2 dimensions will be
established. Apart from providing an interesting link to the 
earlier work on the gauge anomaly in the continuum limit,  
the significance of this result is
that the exact cancellation of the anomaly on the lattice
is reduced to a local cohomology problem which 
appears to be quite tractable.

\section 2. Lattice action and chiral projectors

In the lattice theories studied in this paper the gauge field
couples to a multiplet of left-handed fermions, which transform
according to some unitary representation $R$ of the gauge group $\group$.
We do not impose any restrictions on $R$ or $\group$ at this point
except that $\group$ should be a compact connected Lie group.
As usual the gauge field is represented by link variables
$U(x,\mu)\in\group$, where $x$ runs over all lattice points and 
$\mu=0,\ldots,3$ labels the lattice axes. The lattice is assumed
to be finite with periodic boundary conditions in all directions.

As already mentioned the use of a lattice Dirac operator $D$
satisfying the Ginsparg-Wilson relation is a key element
of the present approach to chiral gauge theories.
While the details of the definition of $D$ are largely irrelevant, 
Neuberger's operator [\ref{NeubergerI}] is an obvious choice
in this context, since it is relatively
simple and has all the required technical properties.
In particular, the locality of the operator
and the differentiability with respect to the gauge field is 
rigorously guaranteed if the gauge field satisfies the bound
$$
  \bigl\|1-R[U(p)]\bigr\|<\epsilon
  \quad\hbox{for all plaquettes $p$,}
  \eqno\enum
$$
where $U(p)$ denotes the product of the link variables around $p$
and $\epsilon$ any fixed positive number less than 
$\frac{1}{30}$ [\ref{Locality}]. 

In the following we shall take it for granted
that the gauge field action restricts
the functional integral to this set of fields.
This can be achieved through a modified
plaquette action, for example [\ref{LuscherAbelian}].
As far as the continuum limit in the weak coupling phase is concerned,
lattice actions of this type should be in the same universality class
as the standard Wilson action,
because the bound (2.1) 
constrains the gauge field fluctuations at the
scale of the cutoff only and does not violate any fundamental
principle such as the locality or 
the gauge invariance of the theory.

Chiral fields may now be defined in a natural way
following the steps previously described in 
refs.~[\ref{HasenfratzNiedermayer}--\ref{OverlapSplit},\ref{LuscherAbelian}].
One first observes that the operator 
$\dirachat=\dirac{5}(1-aD)$ satisfies the relations
$$
  (\dirachat)^{\dagger}=\dirachat,
  \qquad
  (\dirachat)^2=1, 
  \qquad
  D\dirachat=-\dirac{5}D.
  \eqno\enum
$$
The fermion action
$$
  S_{\rm F}=a^4\sum_x \psibar(x)D\psi(x)
  \eqno\enum
$$
thus splits into left- and right-handed parts if the chiral 
projectors for fermion and anti-fermion fields 
are defined through
$$ 
  \hat{P}_{\pm}=\frac{1}{2}(1\pm\dirachat),
  \qquad
  P_{\pm}=\frac{1}{2}(1\pm\dirac{5}),
  \eqno\enum
$$ 
respectively. In particular, by imposing the constraints
$$
  \hat{P}_{-}\psi=\psi,
  \qquad
  \psibar P_{+}=\psibar,
  \eqno\enum
$$
the right-handed components are eliminated 
and one obtains a classical lattice theory where
a multiplet of left-handed Weyl fermions couples to the 
gauge field in a consistent way.

An interesting point to note here is that the space of gauge fields
satisfying the bound (2.1) decomposes into disconnected topological
sectors [\ref{TopA},\ref{TopB}].
In the non-trivial sectors the index of the lattice Dirac operator
[\ref{HasenfratzII},\ref{LuscherSymmetry}]
is in general different from zero and it turns out that 
the dimensions of the spaces of left-handed fermion
and anti-fermion fields are then not the same.
Fermion number violating processes can thus take place,
exactly as expected from the semi-classical approximation 
in continuum chiral gauge theories
[\ref{LuscherAbelian},\ref{Niedermayer}].

\section 3. Fermion integration measure

To complete the definition of the lattice theory,
the functional integration measure for left-handed fermions
needs to be specified.
The principal difficulty here is that  
the constraint (2.5) depends on the gauge field.
This leads to a non-trivial phase ambiguity in the measure 
and eventually gives rise to the gauge anomaly.

To make this clearer let us suppose that $v_j(x)$, $j=1,2,3,\ldots\,$, is 
a basis of complex-valued lattice Dirac fields such that
$$
  \hat{P}_{-}v_j=v_j,
  \qquad
  (v_k,v_j)=\delta_{kj},
  \eqno\enum
$$
the bracket being the obvious scalar product for such fields.
The quantum field may then be expanded according to
$$
  \psi(x)=\sum_j\,v_j(x)c_j,
  \eqno\enum
$$
where the coefficients $c_j$ generate a Grassmann algebra.
They represent the independent degrees of freedom of the field
and an integration measure for left-handed fermion
fields is thus given by
$$
  \rmD[\kern0.5pt\psi\kern0.5pt]=\prod_{j}\,\rmd c_j.
  \eqno\enum
$$
Evidently if we pass to a different basis
$$
  \tilde{v}_j(x)=\sum_l\,v_l(x)(\trans^{-1})_{lj},
  \qquad
  \tilde{c}_j=\sum_l\,\trans_{jl}c_l,
  \eqno\enum
$$
the measure changes by the factor $\det\trans$
which is a pure phase factor since the transformation matrix
$\trans$ is unitary. 

In the following two sets of basis vectors 
$v_j$ and $\tilde{v}_j$
are considered to be equivalent if they are related
to each other through eq.~(3.4) with $\det\trans=1$.
Choosing a fermion integration measure amounts to 
specifying an equivalence class of bases.
A given basis thus represents the associated measure,
but it should not be confused with the measure which is a much simpler object.
In particular, any two fermion measures
coincide up to a gauge field dependent phase factor.
The question of how to fix this phase will occupy
us throughout the rest of this paper. For the time being 
we assume that some particular choice has been made
and proceed with the definition 
of the theory.

In the case of the anti-fermion fields the subspace of left-handed fields
is independent of the gauge field and one can take the same
orthonormal basis $\bar{v}_k(x)$ for all gauge fields.
The ambiguity in the integration measure
$$
  \rmD[\kern0.5pt\psibar\kern0.5pt]=\prod_k\,\rmd \bar{c}_k,
  \qquad
  \psibar(x)=\sum_k\,\bar{c}_k\bar{v}_k(x),
  \eqno\enum
$$
is then only a constant phase factor.
Fermion expectation values of any product $\cal O$ of the basic
fields may now be defined through
$$
  \langle{\cal O}\rangle_{\rm F}=
  \int\rmD[\kern0.5pt\psi\kern0.5pt]\rmD[\kern0.5pt\psibar\kern0.5pt]
  \,{\cal O}\,\rme^{-S_{\rm F}}.
  \eqno\enum
$$
In the non-trivial topological sectors a constant weight factor 
should be included in this formula [\ref{LuscherAbelian}], 
but for brevity this factor is omitted here 
since we shall almost exclusively be concerned 
with the vacuum sector.
The fermion partition function in this sector is
$$
  \langle 1\rangle_{\rm F}=\det M,
  \qquad
  M_{kj}=a^4\sum_{x}\,\bar{v}_k(x)Dv_j(x),
  \eqno\enum
$$
and correlation functions of products of fermion fields may be calculated 
as usual by applying Wick's theorem, the propagator being given by
$$
  \langle\psi(x)\psibar(y)\rangle_{\rm F}=
  \langle 1\rangle_{\rm F}\times
  \hat{P}_{-}S(x,y)P_{+},
  \qquad
  DS(x,y)=a^{-4}\delta_{xy}.
  \eqno\enum
$$
Full normalized expectation values are finally obtained through
$$
  \langle{\cal O}\rangle={1\over{\cal Z}}
  \int \rmD[U]\,\rme^{-S_{\rm G}}\langle{\cal O}\rangle_{\rm F},
  \eqno\enum
$$
where $S_{\rm G}$ denotes the gauge field action,
$\cal Z$ the partition function and $\rmD[U]$ 
the standard integration measure for gauge fields on the lattice.

\section 4. Locality condition

We are now left with the problem to fix the phase of
the fermion integration measure. Evidently this should be done
in such a way that the locality of the theory is preserved
and it also seems reasonable to demand that 
the measure is smoothly dependent on the gauge field.
In this section these conditions are given a precise meaning
and a few key formulae are derived which will later
prove useful when we discuss the gauge anomaly.

The dependence of the fermion measure on the gauge field is best studied
by considering variations\kern2pt%
\footnote{$\dag$}{\footnotefont
Without loss the gauge group $G$ may be assumed to be  
a subgroup of U($n$) for some value of $n$. The generators $T^a$ are then
anti-hermitean matrices and the field components $\eta^a_{\mu}(x)$
are real}
$$
  \delta_{\eta}U(x,\mu)=a\eta_{\mu}(x)U(x,\mu),
  \qquad
  \eta_{\mu}(x)=\eta^a_{\mu}(x)T^a,
  \eqno\enum
$$
of the link field.
Requiring the measure to be smooth means that 
in the neighbourhood of any given gauge field
there exists a differentiable basis $v_j$ of left-handed fields
which represents the measure in the way explained above.
The change $\delta_{\eta}v_j$
of the basis vectors and the linear functional
$$
  \L_{\eta}=i\sum_j\,(v_j,\delta_{\eta}v_j).
  \eqno\enum
$$
are then well-defined. Moreover it is easy to show that 
$\L_{\eta}$ transforms according to 
$$
  \widetilde{\L}_{\eta}=\L_{\eta}-i\delta_{\eta}\ln\det\trans
  \eqno\enum
$$
under basis transformations (3.4).
Equivalent bases thus yield the same linear functional and 
$\L_{\eta}$ is hence a quantity associated with the measure
rather than the basis vectors $v_j$.
Roughly speaking it tells us how the phase of the measure changes
when the gauge field is deformed.

Starting from the definition (3.7), the variation of the effective
action is now easily worked out and one obtains
$$
  \delta_{\eta}\ln\det M=\Tr\{\delta_{\eta}D \hat{P}_{-}D^{-1}P_{+}\}
  -i\L_{\eta}.
  \eqno\enum
$$
The first term in this equation is the naively expected one
while the second arises from the gauge field dependence of the measure.
$\L_{\eta}$ is hence referred to as the measure term in the following.  
Moreover, taking the linearity of $\L_{\eta}$
into account, an associated current $j_{\mu}(x)$ may be defined through
$$
  \L_{\eta}=a^4\sum_x\,\eta^a_{\mu}(x)j_{\mu}^a(x).
  \eqno\enum
$$
As will soon become clear this current plays a central r\^ole in the present
approach to chiral gauge theories. 
In particular, we shall show
in sect.~6 that the measure can be reconstructed from the current
under certain conditions.

Whether a euclidean field theory is local or not
is usually evident from the action.
The situation here is slightly more complicated, because 
the fermion integration measure is not a product of local measures.
An important point to note however is that 
the effective action is the only place where  
the non-trivial structure of the measure shows up. 
In particular, in the vacuum sector 
the fermion integrals $\langle{\cal O}\rangle_{\rm F}$
are equal to the partition function $\langle 1\rangle_{\rm F}$
times a factor which is independent of the measure.

We are thus led to require that the current 
$j_{\mu}(x)$ which is induced by the fermion measure 
is a local expression in the gauge field.
The measure term $\L_{\eta}$ then assumes the form of a local counterterm,
i.e.~the interaction vertices which arise from the
gauge field dependence of the fermion measure are local.
The locality of the theory is thus preserved and
the arbitrariness in the phase of the measure is 
greatly reduced.

\section 5. Gauge anomaly

Another fundamental requirement on the fermion measure 
is that it should not break the gauge symmetry.
In particular, the effective action should be gauge-invariant.
On the lattice the group of gauge transformations 
is connected and it thus suffices to consider infinitesimal gauge 
transformations. One should not conclude from this that 
there are no global anomalies, but as
will become clear later they arise in slightly different ways
than expected from the semi-classical analysis.

Infinitesimal gauge transformations are generated by lattice fields
$\omega(x)$ with values in the Lie algebra of the gauge group. 
The corresponding variations of the link field are obtained by
substituting
$$
  \eta_{\mu}(x)=-\nab{\mu}\omega(x)
  \eqno\enum
$$
in eq.~(4.1), where
the gauge-covariant forward difference operator $\nab{\mu}$ is
given by
$$
  \nab{\mu}\omega(x)={1\over a}\bigl[
  U(x,\mu)\omega(x+a\hat{\mu})U(x,\mu)^{-1}-\omega(x)\bigr]
  \eqno\enum
$$
($\hat{\mu}$ denotes the unit vector in direction $\mu$). 
Taking the transformation behaviour of the Dirac operator into account,
$$
  \delta_{\eta}D=[R(\omega),D],
  \eqno\enum
$$
the terms on the right-hand side of eq.~(4.4) are easily worked out
and for the gauge variation of the effective action 
the result
$$
  \eqalignno{
  &\delta_{\eta}\ln\det M=ia^4\sum_x\,\omega^a(x)
  \left\{\anomaly^a(x)-[\nabstar{\mu}j_{\mu}]^a(x)\right\},
  &\enum\cr
  \noalign{\vskip1.0ex}
  &\anomaly^a(x)={ia\over2}
  \kern1.0pt\tr\{\dirac{5}R(T^a)D(x,x)\},
  &\enum\cr}
$$ 
is thus obtained.
In these equations $\nabstar{\mu}$ 
denotes the gauge-covariant backward difference operator and
$D(x,y)$ the kernel representing the Dirac operator
in position space.
The trace is taken over Dirac and flavour indices only.

We now show that $\anomaly(x)$ 
converges to the covariant gauge anomaly in the classical 
continuum limit. The calculation is practically the same as 
in the case of the axial anomaly which has been studied
in detail in refs.~[\ref{Fujikawa}--\ref{Adams}]. 
One begins by representing the link field through
$$
  U(x,\mu)={\cal P}\exp\left\{
  a\int_0^1 \rmd t\, A_{\mu}(x+(1-t)a\hat{\mu})\right\}, 
  \eqno\enum  
$$
where $\cal P$ implies a path-ordered exponential and
$A_{\mu}(x)$ is an arbitrary smooth gauge potential.
Using the locality and differentiability 
properties of the kernel $D(x,y)$ established 
in ref.~[\ref{Locality}], 
it is then possible to derive an asymptotic expansion
$$
  \anomaly^a(x)\mathrel{\mathop\sim_{a\to0}}
  \sum_{k=0}^{\infty}a^{k-4}{\cal O}^a_k(x).
  \eqno\enum
$$
The fields ${\cal O}_k(x)$ which
occur in this series are traces of $R(T^a)$ times a polynomial of
dimension $k$ in $R[A_{\mu}(x)]$ and its derivatives.  
Moreover they must have the proper transformation behaviour
under the symmetries of the lattice theory.

From eq.~(5.5) it follows that $\anomaly(x)$
is a gauge-covariant pseudo-scalar field which changes sign
when the fermion representation
$R$ is replaced by its complex conjugate.
Taking this into account, it is easy to convince oneself
that all terms ${\cal O}_k(x)$ with dimension $k\leq3$
have to be equal to zero.  
In the continuum limit we are then left with the term
$$
  \anomaly^a(x)=\ca
  \d{abc}\epsilon_{\mu\nu\rho\sigma}
  F^b_{\mu\nu}(x)F^c_{\rho\sigma}(x)+\rmO(a),
  \eqno\enum
$$
where $F_{\mu\nu}(x)$ is the field tensor associated with the
gauge potential and
$$
  \d{abc}=2i
  \kern1pt\tr\bigr\{R(T^a)[R(T^b)R(T^c)+R(T^c)R(T^b)]\bigl\}.
  \eqno\enum
$$
The constant $\ca=-1/128\pi^2$ does not depend on the gauge group
and can be cal\-culated in the U(1) theory with a single
fermion in the fundamental representation [\ref{Fujikawa}--\ref{ChiuHsieh}].
Since the gauge anomaly coincides with the axial anomaly in this case,
the number may also be inferred from the index theorem
[\ref{HasenfratzII},\ref{LuscherSymmetry},\ref{LuscherAnomaly}].

Returning to the question posed at the beginning of this section,
the results obtained above show that the effective action is 
gauge-invariant if (and only if)
$$
  [\nabstar{\mu}j_{\mu}]^a(x)=\anomaly^a(x).
  \eqno\enum
$$
In other words, the phase of the fermion measure should be chosen so
that the associated current satisfies this equation.
Together with eq.~(4.4) 
the gauge invariance of the effective action moreover implies that
$j_{\mu}(x)$ has to be gauge-covariant.

As is well-known one cannot have both, locality and 
gauge invariance, unless the anomaly cancellation condition
$$
  \d{abc}=0
  \eqno\enum
$$
is fulfilled. There is more than one way to prove this in the present
framework, a quick argument being that an expansion similar to
eq.~(5.7) must exist in the case of the current $j_{\mu}(x)$ too,
since it is required to be local and smoothly dependent on the gauge field.
In the continuum limit the anomaly is hence equal to the
divergence of a covariant local current which is a polynomial of
dimension $3$ in the gauge potential and its derivatives. This is only
possible if the anomaly vanishes at $a=0$, i.e.~if eq.~(5.11) holds.

\section 6. Integrability condition

So far we have assumed that 
the current $j_{\mu}(x)$ is obtained from a given fermion measure
through eqs.~(4.2) and (4.5).
We now show that any prescribed current satisfying 
a certain integrability condition arises from a measure in this way.
The relation between the measure and the current is thus invertible
and one may adopt the point of view that the latter 
is the fundamental object.

To derive the integrability condition
we need to study the change of phase of the fermion measure
along smooth curves 
$$
  U_t(x,\mu),
  \qquad
  0\leq t\leq 1,
  \eqno\enum
$$
in the space of gauge fields.
As discussed in sect.~4, the measure term $\L_{\eta}$ tells us 
how the phase varies when the gauge field is deformed
in a particular direction.
The total change of phase along any given curve 
is thus given by the Wilson line
$$
  \Wline=\exp\left\{i\int_0^1\rmd t\,\L_{\eta}\right\},
  \qquad
  a\eta_{\mu}(x)=
  \partial_tU_t(x,\mu)U_t(x,\mu)^{-1}.
  \eqno\enum
$$
In general this phase is non-integrable, i.e.~the Wilson lines
around closed curves
are not necessarily equal to $1$.
To work this out we introduce the projector
$$
  P_t=\hat{P}_{-}\bigr|_{U=U_t}
  \eqno\enum
$$
and define a unitary operator $Q_t$ through the differential equation
$$
  \partial_tQ_t=\left[\partial_t P_t,P_t\right]Q_t,
  \qquad
  Q_0=1.
  \eqno\enum
$$
It is easy to prove that 
$$
  P_tQ_t=Q_tP_0
  \eqno\enum
$$
and $Q_t$ thus transports the projector $P_t$ along the curve.
In a few lines (appendix~A) it is then possible to 
establish the identity
$$
  W=\det\{1-P_0+P_0Q_1\} 
  \eqno\enum
$$
for all closed curves.
The Wilson loops are hence the same for all
fermion measures. In other words, they represent 
a geometrical invariant of the measure.

Let us now assume that $j_{\mu}^a(x)$ is an arbitrary
current depending smoothly on the gauge field.
$\L_{\eta}$ and the associated Wilson lines $W$ 
may then be defined through eqs.~(4.5) and (6.2) respectively.
Evidently, for the current to arise from a fermion measure,
a necessary condition is that eq.~(6.6) holds for all closed loops.
This is in fact also a sufficient condition for the 
existence of a such a measure.
Moreover, in each topological sector 
this measure is uniquely determined, 
up to a constant phase factor, and smooth.

To prove these statements we consider a definite topological sector
and choose an arbitrary reference field $U_0$ in this sector.
Any other field $U$ in the same sector
can then be reached through a smooth curve $U_t$ 
such that $U_1=U$. If we define $Q_t$ through eq.~(6.4) as before, 
a basis of left-handed fields at the point $U$ is given by
$$
  v_j=\cases{Q_1w_1\Wline^{-1} & if $j=1$, \cr
             \noalign{\vskip1ex}
             Q_1w_j       & otherwise, \cr}
  \eqno\enum
$$
where $w_j$ denotes a fixed basis at the reference point
and $W$ the Wilson line (6.2) computed from the given current.
This basis is path-dependent, but the associated measure is not,
because any two curves $U_t$ and $\widetilde{U}_t$ 
form a closed loop and the integrability
condition (6.6) then implies that the unitary transformation 
relating the basis vectors $v_j$ and $\tilde{v}_j$ has determinant $1$.
Taking this into account,
it is easy to show that the fermion measure 
defined by the basis (6.7)
has all the properties mentioned above.

The construction of chiral gauge theories on the lattice 
is thus reduced to the problem of finding
a local current which fulfils the integrability condition (6.6) 
and the requirement of gauge invariance.
Once this is achieved, the theory is completely specified
up to a constant phase factor in each topological sector.
Note that it suffices to define the current for all gauge
fields satisfying the bound (2.1) since only these contribute
to the functional integral.

\section 7. Fermion determinant

Using the results obtained in the preceding section, we are now
in a position to derive a closed expression for the fermion determinant 
in terms of the Dirac operator and the current $j_{\mu}(x)$.
We shall then be able to make contact with Kaplan's 
approach to chiral gauge theories [\ref{Kaplan}] and the earlier
work of Alvarez-Gaum\'e et al.~[\ref{AlvarezEtAl},\ref{NielsBohr}]
and Ball and Osborn [\ref{BallOsborn},\ref{Ball}]
on the effective action in the continuum theory.

Suppose $U_0$ is an arbitrary reference field in the vacuum sector 
and let $w_j$ be a basis of left-handed fermion fields
at this point. As explained above, the basis (6.7) then represents
the fermion measure at any other point $U$ in the vacuum sector,
up to a constant phase factor.
If we insert this basis in eq.~(3.7), the formula
$$
  \det M 
  \det M_{\kern0.3pt\smash{\raise0.8pt\hbox{$\scriptstyle 0$}}}
        ^{\smash{\kern0.3pt\raise1.5pt\hbox{$\scriptstyle\dagger$}}}
  =
  \det\bigl\{1-P_{+}+P_{+}DQ_1
  D_{\kern0.3pt\smash{\raise0.8pt\hbox{$\scriptstyle 0$}}}
        ^{\smash{\kern0.3pt\raise1.5pt\hbox{$\scriptstyle\dagger$}}}
  \bigr\}\,\Wline^{-1}
  \eqno\enum
$$
is obtained, where $D_0$ and $M_0$ denote the Dirac operator and 
fermion matrix at the reference point\kern2pt%
\footnote{$\dag$}{\footnotefont
Another expression for the effective action, involving the Dirac operator
and the current $j_{\mu}(x)$ only, may be obtained by integrating
eq.~(4.4) along any particular path.
The idea has recently been pursued by Suzuki [\ref{SuzukiAction}]
in abelian chiral gauge theories
}.
All other notations are as in eq.~(6.7).
To fully understand this result
the following remarks may be helpful.

\vskip0.8ex\noindent
(a)~The integrability condition guarantees that 
the right-hand side of the equation does not depend on the curve $U_t$
which has been chosen to connect $U=U_1$ with $U_0$.
In other words, the path-dependence of the determinant 
and the Wilson line $\Wline$ precisely cancel each other.
Note incidentally that the constant phase ambiguity of the 
measure drops out in the product of determinants. 

\vskip0.8ex\noindent
(b)~Using eqs.~(2.2) and (6.5) it is easy to check that 
$DQ_1D_{\kern0.3pt\smash{\raise0.8pt\hbox{$\scriptstyle 0$}}}
        ^{\smash{\kern0.3pt\raise1.0pt\hbox{$\scriptstyle\dagger$}}}$
commutes with $\dirac{5}$. 
The determinant of this operator in
the subspace of left-handed anti-fermion fields
coincides with the determinant 
on the right-hand side of eq.~(7.1)
and the chiral nature of the expression is thus evident.

\vskip0.8ex\noindent
(c)~In sect.~8 we shall show that the current 
$j_{\mu}(x)$ vanishes in the 
classical continuum limit if the fermion multiplet
is anomaly-free. The Wilson line $\Wline$
consequently does not contribute to the fermion determinant in this limit.

\vskip0.8ex\noindent
(d)~Concerning the operator $Q_t$ we note that the differential
equation (6.4) may be rewritten in the form
$$
  \partial_tQ_t=\frac{1}{2}a
  \left[\dirac{5}\partial_tD_t,P_t\right]Q_t,
  \qquad
  D_t=\left.D\right|_{U=U_t}.
  \eqno\enum
$$
Close to the classical continuum limit, and when acting on 
fermion fields with frequencies far below the lattice cutoff,
the operator is hence equal to $1$ up to terms of order $a$.
In particular, in eq.~(7.1) the operator $Q_1$
only affects the contribution of the high-frequency modes 
and it may, therefore, be regarded as part of the lattice
regularization prescription for the chiral determinant.

\vskip0.8ex\noindent
(e)~So far the reference field $U_0$ has been assumed
to be fixed and the factor 
$\det M_{\kern0.3pt\smash{\raise0.8pt\hbox{$\scriptstyle 0$}}}
        ^{\smash{\kern0.3pt\raise1.0pt\hbox{$\scriptstyle\dagger$}}}$
on the left-hand side of eq.~(7.1) is then just a constant. 
Since $U_0$ and $U_1$ are inter\-changeable in this equation,
another option is to interpret $U_0$ as a second gauge field and  
$\det M_{\kern0.3pt\smash{\raise0.8pt\hbox{$\scriptstyle 0$}}}
        ^{\smash{\kern0.3pt\raise1.0pt\hbox{$\scriptstyle\dagger$}}}$
as the determinant arising from
a multiplet of right-handed fermions.
In the present framework the formulation of  
such left-right symmetric chiral gauge theories
thus appears to be particularly natural.

\vskip1.0ex
Having clarified the structure of eq.~(7.1),
we now briefly discuss how the formula relates to Kaplan's 
approach to chiral gauge theories [\ref{Kaplan}].
In the version proposed by Shamir [\ref{Shamir}],
this approach starts from a gauge theory in 4+1 dimensions with 
a multiplet of massive Dirac fermions,
where the additional coordinate is assumed to range between $0$
and $T$ with Dirichlet boundary conditions 
on the gauge and fermion fields.
The reduction to 4 dimensions is then achieved by noting that 
the Dirac operator admits chiral surface modes
whose interactions at large $T$
are described by an effective chiral gauge theory.

Contact with the present framework can now be made if 
we identify the fifth coordinate, scaled to the range $[0,1]$, 
with the parameter $t$ of the path $U_t$.
The path thus becomes a gauge field in 4+1 dimensions
with boundary values $U_0$ and $U_1$.
In particular, we can compare the fermion determinant in 4+1 dimensions
with the determinant on the right-hand side of eq.~(7.1) and 
it is then conceivable that they agree in the limit
where the lattice spacing in the fifth dimension is sent to 
$0$ and $T$ to infinity.
Preliminary studies suggest that 
this is indeed what happens if the 
lattice Dirac operator in 4+1 dimensions is chosen appropriately,
but the details are complicated and will not be presented here.
It is interesting to note, however,
that from the point of view of the higher-dimensional theory,
the Wilson line $\Wline$ in eq.~(7.1) amounts to 
adding a local counterterm to the gauge field action.
The term cancels the dependence of the fermion determinant
on the gauge field in the interior of the space-time volume
and thus allows one to reduce the theory to 4 dimensions,
the dynamically relevant degrees of freedom being the boundary
values $U_0$ and $U_1$.

In the continuum limit the phase of the fermion determinant
is known to be proportional to 
the $\eta$-invariant of the Dirac operator
in 4+1 dimensions [\ref{AlvarezEtAl}--\ref{Ball}]. 
One considers the massless Dirac operator in this case
and uses Pauli-Villars regularization or analytic continuation 
methods to define the determinant, but otherwise
the setup is the same as the one described above.
The $\eta$-invariant, Kaplan's approach and the results obtained here
are hence closely related to each other.
As discussed by Kaplan and Schmaltz [\ref{KaplanSchmaltz}],
the formula for the effective
action of refs.~[\ref{AlvarezEtAl}--\ref{Ball}]
may in fact be directly derived from 
the fermion integral in 4+1 dimensions.

\section 8. Classical continuum limit

To complete the construction of the lattice theory
we still need to prove that there exists 
a local current $j_{\mu}(x)$ satisfying 
the requirement of gauge invariance and 
the integrability condition.
The aim in the following lines is to 
determine the general solution of this problem in the classical
continuum limit.
Along the way an important simplification is achieved by considering
the integrability condition in its differential form.
Global anomalies and the relation between this equation 
and the gauge anomaly
are further topics which will be addressed.

The differential form of the integrability condition,
$$
  \delta_{\eta}\L_{\zeta}-\delta_{\zeta}\L_{\eta}
  +a\L_{[\eta,\zeta]}=
  i\kern1pt\Tr\bigl\{\hat{P}_{-}[\delta_{\eta}\hat{P}_{-},
                         \delta_{\zeta}\hat{P}_{-}]\bigr\},
  \eqno\enum
$$
is obtained by computing the variation of the 
Wilson loop and the determinant in eq.~(6.6) under 
infinitesimal deformations of the loop $U_t$.
The equation may also be derived in a more direct way, starting
from the representation (4.2) of the measure term
and making use of the identity
\footnote{$\dag$}{\footnotefont
In eqs.~(8.1) and (8.2) 
the variations $\eta$ and $\zeta$ are assumed to be 
independent of the gauge field.
Further terms proportional to $\delta_{\eta}\zeta$ and 
$\delta_{\zeta}\eta$ have to be included if this is not the case}
$$
  \delta_{\eta}\delta_{\zeta}-\delta_{\zeta}\delta_{\eta}
  +a\delta_{[\eta,\zeta]}=0
  \eqno\enum
$$
and the defining properties of the basis vectors $v_j$.

An important point to note is that
the integrability condition is a slightly 
stronger constraint than its differential form.
To make this completely clear, let us assume that 
$\L_{\eta}$ is an arbitrary linear functional
satisfying eq.~(8.1), for all gauge fields
com\-plying with the bound (2.1)
and all vector fields $\eta_{\mu}(x)$ and $\zeta_{\mu}(x)$.
The Wilson loop associated with any closed curve $U_t$
of fields is then given by
$$
  W=h\det\{1-P_0+P_0Q_1\},
  \eqno\enum
$$
where $h$ is invariant under continuous deformations of the curve.
Evidently $h$ is equal to 1 for all contractible loops,
but in general this need not be so 
and eq.~(6.6) thus imposes an additional constraint
on $\L_{\eta}$ if there are topologically non-trivial loops.
Presumably the global anomalies discovered
by Witten [\ref{Witten}] are related to this observation
since they arise from certain non-contractible loops
in the space of gauge orbits [\ref{ElitzurNair}].
Further studies are however required before a definite answer
to this question can be given
[\ref{BaerCampos}]. 

In the classical continuum limit the differential form
of the integrability condition reduces to 
a simple equation. To show this we follow the steps
previously described in sect.~5,
i.e.~we insert the representation (5.6) for the link variables 
and assume that $\eta_{\mu}(x)$ and $\zeta_{\mu}(x)$ are 
restrictions to the lattice of some differentiable vector fields.
The locality properties of the Dirac operator [\ref{Locality}] then imply
$$
  i\kern1pt\Tr\bigl\{\hat{P}_{-}[\delta_{\eta}\hat{P}_{-},
                         \delta_{\zeta}\hat{P}_{-}]\bigr\}
  \mathrel{\mathop\sim_{a\to0}}
  \sum_{k=0}^{\infty}a^{k-4}\int\rmd^4x\,{\cal O}_k(x),
  \eqno\enum
$$ 
where the fields ${\cal O}_k(x)$
are traces of polynomials in $R[A_{\mu}(x)]$,
$R[\eta_{\mu}(x)]$, $R[\zeta_{\mu}(x)]$ 
and their derivatives.
Taking the symmetries of the expression into account,
this leads to the result
$$
  i\kern1pt\Tr\bigl\{\hat{P}_{-}[\delta_{\eta}\hat{P}_{-},
                         \delta_{\zeta}\hat{P}_{-}]\bigr\}
  =\ci
  \int\rmd^4x\,
  \d{abc}\epsilon_{\mu\nu\rho\sigma}
  \eta^a_{\mu}(x)\zeta^b_{\nu}(x)F^c_{\rho\sigma}(x)+\rmO(a),
  \eqno\enum
$$
the notations being the same as in sect.~5. 

The proportionality constant $\ci$ in this equation
is related to the coefficient $\ca$ of the anomaly. 
To work this out we consider
a gauge variation $\eta_{\mu}(x)=-\nab{\mu}\omega(x)$
and note that
$$
  \delta_{\eta}\hat{P}_{-}=[R(\omega),\hat{P}_{-}],
  \qquad
  \hat{P}_{-}\delta_{\zeta}\hat{P}_{-}\hat{P}_{-}=0.
  \eqno\enum
$$
It is then straightforward to establish the identity
$$
  i\kern1pt\Tr\bigl\{\hat{P}_{-}[\delta_{\eta}\hat{P}_{-},
                         \delta_{\zeta}\hat{P}_{-}]\bigr\}
  =-a^4\sum_x\omega^a(x)\delta_{\zeta}\anomaly^a(x)
  \eqno\enum
$$
and after substituting the asymptotic forms, eqs.~(5.8) and (8.5),
the relation
$$
  \ci=-4\ca=1/32\pi^2
  \eqno\enum
$$
is thus obtained. 
In passing we remark that the anomalous conservation law (5.10)
is consistent with the integrability condition (8.1) in the 
sense that the combination of these equations does not lead to further
constraints on the current $j_{\mu}(x)$ apart from the fact  
that it should transform covariantly
under gauge transformations.

An important conclusion which can be drawn at this point is that
$j_{\mu}(x)=0$ is an acceptable choice of the current 
in the classical continuum limit if the fermion multiplet is 
anomaly-free. Both, the requirement of gauge invariance and 
the integrability condition in its differential form, are then satisfied. 
There is in fact no other sensible solution since 
it is impossible to construct a gauge-covariant
polynomial of dimension 3 in the gauge potential $A_{\mu}(x)$ and 
its derivatives which transforms as an axial vector current.
As far as the classical continuum limit is concerned,
the theory is thus completely specified up to a constant phase factor in each
topological sector.

\section 9. Equivalent cohomology problem in 4+2 dimensions 

Most gauge field configurations which contribute to
the functional integral 
are not as smooth as those
considered in the classical continuum limit
and a general strategy to determine the current $j_{\mu}(x)$
thus needs to be developed if one is interested in 
constructing the complete theory.
As a first step in this direction, we here show 
that the anomalous conservation law (5.10) and the
integrability condition (8.1) can be mapped to a local cohomology problem
whose solution is known to
all orders in the lattice spacing. 

To explain which type of cohomology problem we are heading to, 
let us consider the pure gauge theory on $\rz^n$ with gauge group $\group$
and suppose $q(z)$ is a gauge-invariant polynomial in the gauge 
potential $A_{\alpha}(z)$ and its derivatives.
Such fields 
are called {\it topological}\/ if
$$
  \int\rmd^nz\,\delta q(z)=0
  \eqno\enum
$$
for any variation $\delta A_{\alpha}(z)$ of the gauge potential with 
compact support. 
Using the Stora-Zumino descent equations [\ref{StoraI}--\ref{Zumino}],
it is possible to prove that all topological fields are of the form
$$
  q(z)=c(z)+\partial_{\alpha}k_{\alpha}(z),
  \eqno\enum
$$
where $c(z)$ is a linear combinations of Chern monomials
$$
  c_{\alpha_1\ldots\alpha_{2m}}t^{a_1\ldots a_m}
  F^{a_1}_{\alpha_1\alpha_2}(z)\ldots F^{a_m}_{\alpha_{2m-1}\alpha_{2m}}(z)
  \eqno\enum
$$
and $k_{\alpha}(z)$ a gauge-invariant 
local current [\ref{BrandtEtAl}--\ref{Dragon}].
The tensor $c_{\alpha_1\ldots\alpha_{2m}}$ in this expression has to 
be totally anti-symmetric and $t^{a_1\ldots a_m}$
should be invariant under the adjoint action of the gauge group.

The classification of topological fields modulo divergence terms
is a particular case of a local cohomology problem in which the 
gauge symmetry plays an important r\^ole.
From this point of view the Chern monomials represent the 
non-trivial cohomology classes. 
Depending on the gauge group, a basis of linearly independent 
Chern monomials is usually not difficult to find
(see ref.~[\ref{Gilkey}] for example).

Returning to the lattice, our aim in the following paragraphs is to 
construct a topological field in 4+2 dimensions
whose cohomology class is trivial if (and only if)
there exists a local current $j_{\mu}(x)$ with the required properties.
The added dimensions are continuous, 
i.e.~we are concerned with lattice gauge fields
$$
  U(z,\mu)\in\group,
  \qquad
  z=(x,t,s),
  \qquad
  \mu=0,\ldots,3,
  \eqno\enum
$$
which depend on two additional real coordinates $t$ and $s$.
We also introduce gauge potentials $A_t(z)$, $A_s(z)$ along these directions
and define the associated field tensor through
$$
  F_{ts}(z)=\partial_tA_s(z)-\partial_sA_t(z)+[A_t(z),A_s(z)].
  \eqno\enum
$$
Under arbitrary gauge transformations in 4+2 dimensions,
the covariant derivative
$$
  \D_rU(z,\mu)=\partial_rU(z,\mu)+A_r(z)U(z,\mu)-U(z,\mu)A_r(z+a\hat{\mu})
  \eqno\enum
$$
then transforms in the same way as $U(z,\mu)$
and a similar statement also applies to the derivatives 
$$
  \D_r\hat{P}_{-}=\partial_r\hat{P}_{-}+[R(A_r),\hat{P}_{-}]
  \eqno\enum
$$
of the projector $\hat{P}_{-}$
(here and below the index $r$ stands for $t$ or $s$).

We now consider the field
$$
  \eqalignno{
  &q(z)=-i\kern1pt\tr\Bigl\{
  \Bigl[\frac{1}{4}\dirachat[\D_t\hat{P}_{-},\D_s\hat{P}_{-}]+
  \frac{1}{4}[\D_t\hat{P}_{-},\D_s\hat{P}_{-}]\dirachat
  &\cr
  \noalign{\vskip1.5ex}
  &\kern5.1em+\frac{1}{2}R(F_{ts})\dirachat\Bigr](x,x)\Bigr\},
  &\enum\cr}
$$
where $\left[\ldots\right](x,y)$ denotes the kernel
representing the operator enclosed in the square bracket,
at fixed $t$ and $s$, in the same way as $D(x,y)$ represents 
the Dirac operator. 
The trace is taken over the Dirac and flavour indices only
and $q(z)$ is thus a gauge-invariant local field
in 4+2 dimensions. 
It is also not difficult to check (appendix B) that 
$q(z)$ satisfies 
$$
  a^4\sum_x\int\rmd t\,\rmd s\,
  \delta q(z)=0
  \eqno\enum
$$
for all local variations of the link variables $U(z,\mu)$ 
and the potential $A_r(z)$, i.e.~it is a topological field.

By definition $q(z)$ is in the trivial cohomology class if it 
is equal to the divergence of a gauge-invariant local current.
We now show that this implies
the existence of a local current
$j_{\mu}(x)$ in 4 dimensions satisfying the anomalous conservation law (5.10)
and the integrability condition (8.1).
The converse is also true, but we shall not prove this here.

So let us suppose that 
$$
  q(z)=\drvstar{\mu}k_{\mu}(z)+\partial_tk_s(z)-\partial_sk_t(z),
  \eqno\enum
$$
where $k_{\mu}(z)$, $k_t(z)$ and $k_s(z)$ 
are gauge-invariant polynomials in
$$
  \partial^n_t\partial^m_sU(z,\mu),
  \qquad
  \partial^n_t\partial^m_sA_r(z),
  \qquad
  n+m\geq1,
  \eqno\enum
$$
with coefficients that are local fields on the lattice
depending on $U(z,\mu)$ and $A_r(z)$.
In eq.~(9.10) the symbol $\drvstar{\mu}$
denotes the backward difference operator 
and the last two terms have been written in the form of a curl for 
reasons to become clear below.

Under scale transformations of the coordinates $t$ and $s$,
the monomials contributing to $k_{\mu}(z)$, $k_t(z)$ and $k_s(z)$
transform homogeneously if $A_r(z)$ is transformed
in the usual way. It is then immediately clear that
all terms on the right-hand side of eq.~(9.10)
with scale dimensions different from those of the left-hand side
have to cancel.
Without loss we may, therefore, assume that the $(t,s)$-dimensions
of $k_{\mu}(z)$, $k_t(z)$ and $k_s(z)$ are 
$(1,1)$, $(1,0)$ and $(0,1)$ respectively. 
Taking the gauge symmetry into account, this implies
$$
  \eqalignno{
  &k_r(z)=a^4\sum_y\lambda^a_{r,\mu}(w)K^a_{r,\mu}(w,z),
  \qquad 
  w=(y,t,s),
  &\enum\cr
  \noalign{\vskip1ex}
  &a\lambda_{r,\mu}(w)=\D_rU(w,\mu)U(w,\mu)^{-1},
  &\enum\cr}
$$
where $K_{r,\mu}(w,z)$ is a gauge-covariant 
local expression in the link variables.

After summing over all
lattice points, eq.~(9.10) thus assumes the form
$$
  \eqalignno{
  &a^4\sum_y\bigl\{\partial_t\bigl[\lambda^a_{s,\mu}(w)j^a_{s,\mu}(w)\bigr]
  -\partial_s\bigl[\lambda^a_{t,\mu}(w)j^a_{t,\mu}(w)\bigr]\bigr\}
  &\cr
  \noalign{\vskip0.5ex}
  &\phantom{j^a_{r,\mu}(w)}
  =i\kern1pt\Tr\bigl\{\hat{P}_{-}[\D_t\hat{P}_{-},\D_s\hat{P}_{-}]
  -\frac{1}{2}R(F_{ts})\dirachat\bigr\},
  &\enum\cr
  \noalign{\vskip2.5ex}
  &j^a_{r,\mu}(w)=a^4\sum_x K^a_{r,\mu}(w,z).
  &\enum}
$$
Evidently $j_{r,\mu}(w)$ is a gauge-covariant local current
depending on the link variables at the given values
of $t$ and $s$, but not on their
derivatives with respect to these coordinates.
Collecting all terms in eq.~(9.14) proportional to
$\partial_t\partial_s U(w,\mu)$, we thus conclude that 
$$
  j_{t,\mu}(w)=j_{s,\mu}(w)\equiv j_{\mu}(w).
  \eqno\enum
$$
Note that the $t,s$-dependence of $j_{\mu}(w)$
arises through the link variables only, i.e.~the current may
be considered to be a local field in 4 dimensions
which has been extended to 4+2 dimensions by letting the gauge field
depend on $t$ and $s$.

It is now straightforward to show that this current
has all the required properties.
The anomalous conservation law (5.10), for example,
follows from eq.~(9.14) by choosing the link variables to be independent
of $t$ and $s$. 
We may also set $A_r(z)=0$ in this equation
and in a few lines one then finds that the integrability condition
(8.1) is fulfilled.
This proves that the triviality of the cohomology class of $q(z)$
implies the existence a local current satisfying eqs.~(5.10) and (8.1).

In the classical continuum limit
the cohomology class of $q(z)$ is easily determined.
As in the case of the anomaly discussed
in sect.~5, the field may be expanded in a power
series in $a$, the leading term being given by
$$
  q(z)=\frac{1}{6}c_1\d{abc}\epsilon_{\alpha_1\ldots\alpha_6}
  F^a_{\alpha_1\alpha_2}(z)F^b_{\alpha_3\alpha_4}(z)F^c_{\alpha_5\alpha_6}(z)
  +\rmO(a).
  \eqno\enum
$$
The obvious notations for gauge fields in $n=6$ dimensions are being
used in this equation, with space-time indices running from 0 to 5. If
the fermion multiplet is anomaly-free, the expression on the
right-hand side (which is equal to $2\pi$ times the Chern character
[\ref{Gilkey}]) vanishes and in this case
$q(z)$ has trivial cohomology to lowest order in $a$.  
The same is true to any order in $a$, 
because the local fields which one generates
at the higher orders of the expansion are all topological.  
Recalling the theorem quoted at the beginning
of this section, this implies that they are of the form (9.2) with
$c(z)=0$ since there are no Chern monomials with scale dimension
greater than the space-time dimension.

The existence of a local current
satisfying eqs.~(5.10) and (8.1) is thus guaranteed to all orders in $a$.
Moreover all what is needed to extend this result
to any fixed value of the lattice spacing
is the classification of all topological fields in 4+2 dimensions.
Presumably the non-trivial cohomology classes on the lattice 
are in one-to-one correspondence
with the linearly independent Chern monomials.
At least this is so in the
abelian case [\ref{LuscherAnomaly}].

\section 10. Concluding remarks

When studying chiral gauge theories 
one is often led to consider gauge and fermion fields in higher dimensions.
It may well be that this is just a matter of mathematical convenience.
On the other hand, the experience should perhaps be taken as an indication 
that chiral gauge theories are merely effective descriptions of the 
low-energy modes of a more fundamental theory in 4+1 or 4+2 dimensions.
The approach of Kaplan and Shamir [\ref{Kaplan},\ref{Shamir}]
provides a concrete model for this and it would be important 
to work out its relation to the framework presented in this paper
in full detail, following the lines sketched in sect.~7.

In the continuum limit the gauge anomaly cancels
if the tensor $\d{abc}$ vanishes.
The same is presumably true on the lattice,
but a complete proof of this has only been given 
in abelian theories so far [\ref{LuscherAbelian}]. 
For non-abelian gauge groups the current status is that the 
anomaly cancellation has been established to all orders
of an expansion in powers of the lattice spacing. 
Moreover, as explained in sect.~9, 
the problem has been reduced to classifying the topological fields
in 4+2 dimensions, which does not seem to be an impossible task.

Global anomalies are a separate issue which requires control over
the first homotopy group of the space of lattice gauge fields
satisfying the bound (2.1). 
One may be able to achieve this by noting that such fields are continuous
on the scale of the lattice spacing up to gauge transformations.
The topology of the space of gauge orbits is hence expected to be 
essentially the same as in the continuum theory.

\vskip1ex
I am grateful to Raymond Stora for guiding me through the 
literature on local cohomology in gauge theories.
Thanks also go to Oliver B\"ar and
Isabel Campos for helpful discussions on global anomalies 
and to Peter Weisz for a critical reading of a first draft of this paper.

\appendix A

To prove eq.~(6.6) we choose a differentiable basis $v_j$ of left-handed 
fields representing the fermion measure along the curve
such that $v_j|_{t=1}=v_j|_{t=0}$ (this is always possible
if the measure is smooth). 
The measure term in eq.~(6.2) is then given by
$$
  \L_{\eta}=i\sum_j\,(v_j,\partial_t v_j).
  \eqno\enum
$$
Taking the properties of the operator $Q_t$ into account, we have
$$
  v_j=Q_t\sum_lv_l|_{t=0}(\vartrans^{-1})_{lj},
  \eqno\enum
$$
where $\vartrans$ is some unitary transformation matrix satisfying 
$\vartrans|_{t=0}=1$. When inserted in eq.~(A.1) this yields
$$
  \L_{\eta}=-i\partial_t \ln\det\vartrans
  \eqno\enum
$$
and the Wilson loop $\Wline$ is thus equal to $\det\vartrans|_{t=1}$.
At $t=1$ the matrix $\vartrans$ represents the action of $Q_1$
in the subspace of left-handed fields. In particular, its determinant
coincides with the determinant of $Q_1$ in this subspace, i.e.~with the
right-hand side of eq.~(6.6).

\appendix B

Starting from the definition (9.8)
it is straightforward to show that 
$$
  a^4\sum_x q(z)=
  i\kern1pt\Tr\bigl\{\hat{P}_{-}
  [\partial_t\hat{P}_{-},\partial_s\hat{P}_{-}]
  -\frac{1}{2}\partial_t[R(A_s)\dirachat]
  +\frac{1}{2}\partial_s[R(A_t)\dirachat]
  \bigr\}.
  \eqno\enum
$$
As a consequence we have
$$
  a^4\sum_x\int\rmd t\,\rmd s\,
  \delta q(z)=\int\rmd t\,\rmd s\,
  \delta\Bigl\{i\kern1pt\Tr\bigl\{\hat{P}_{-}
  [\partial_t\hat{P}_{-},\partial_s\hat{P}_{-}]\bigr\}\Bigr\}
  \eqno\enum
$$
for any local deformation of the gauge field.
To evaluate the right-hand side of this equation,
we make use of the identity 
$$
  \Tr\bigl\{
  \delta\hat{P}_{-}\kern1pt\partial_t\hat{P}_{-}\kern1pt\partial_s\hat{P}_{-}
  \bigr\}=0,
  \eqno\enum
$$
which may be established by inserting $(\dirachat)^2=1$ and noting that
$\dirachat$ anti-commutes with the derivatives of the projector.
One then finds that the integrand is given by
$$
  \partial_t\Bigl\{i\kern1pt\Tr\bigl\{\hat{P}_{-}
  [\delta\hat{P}_{-},\partial_s\hat{P}_{-}]\bigr\}\Bigr\}-
  \partial_s\Bigl\{i\kern1pt\Tr\bigl\{\hat{P}_{-}
  [\delta\hat{P}_{-},\partial_t\hat{P}_{-}]\bigr\}\Bigr\}
  \eqno\enum
$$
and after integrating over $t$ and $s$ one gets zero because
the variation of the gauge field is compactly supported.


\ninepoint

\beginbibliography


\bibitem{GinspargWilson}
P. H. Ginsparg and K. G. Wilson,
Phys. Rev. D25 (1982) 2649

\bibitem{HasenfratzI}
P. Hasenfratz,
Nucl. Phys. B (Proc. Suppl.) 63A-C (1998) 53;
Nucl. Phys. B525 (1998) 401 

\bibitem{HasenfratzII}
P. Hasenfratz, V. Laliena and F. Niedermayer,
Phys. Lett. B427 (1998) 125

\bibitem{NeubergerI}
H. Neuberger,
Phys. Lett. B417 (1998) 141;
{\it ibid}\/ B427 (1998) 353

\bibitem{LuscherSymmetry}
M. L\"uscher,
Phys. Lett. B428 (1998) 342


\bibitem{LuscherAbelian}
M. L\"uscher,
Abelian chiral gauge theories on the lattice with 
exact gauge invariance,
hep-lat/9811032, to appear in Nucl. Phys. B


\bibitem{LuscherAnomaly}
M. L\"uscher,
Nucl. Phys. B538 (1999) 515


\bibitem{StoraI}
R. Stora,
Continuum gauge theories,
in: New developments in quantum field theory and statistical mechanics
(Carg\`ese 1976),
eds. M. L\'evy and P. Mitter (Plenum Press, New York, 1977)

\bibitem{StoraII}
R. Stora,
Algebraic structure and topological origin of anomalies,
in: Progress in gauge field theory 
(Carg\`ese 1983),
eds. G. `t Hooft et al. (Plenum Press, New York, 1984)

\bibitem{Zumino}
B. Zumino,
Chiral anomalies and differential geometry,
in: Relativity, groups and topology 
(Les Houches 1983), 
eds. B. S. DeWitt and R. Stora
(North-Holland, Amsterdam, 1984)


\bibitem{AlvarezGinsparg}
L. Alvarez-Gaum\'e and P. Ginsparg,
Nucl. Phys. B243 (1984) 449


\bibitem{EriceLectures}
L. Alvarez-Gaum\'e,
An introduction to anomalies, 
in: Fundamental problems of gauge field theory (Erice 1985),
eds. G. Velo and A. S. Wightman (Plenum Press, New York, 1986)

\bibitem{Bertlmann}
R. A. Bertlmann, Anomalies in quantum field theory (Oxford University
Press, Oxford, 1996)


\bibitem{Locality}
P. Hern\'andez, K. Jansen and M. L\"uscher,
Locality properties of Neuberger's lattice Dirac operator,
hep-lat/9808010, to appear in Nucl. Phys. B


\bibitem{HasenfratzNiedermayer}
P. Hasenfratz and F. Niedermayer,
private communication (February 1998)

\bibitem{Niedermayer}
F. Niedermayer,
Exact chiral symmetry, topological charge and related topics,
Talk given at the International Symposium on Lattice Field Theory,
Boulder 1998, \hfill\break
hep-lat/9810026

\bibitem{OverlapSplit}
R. Narayanan,
Phys. Rev. D58 (1998) 97501


\bibitem{TopA}
M. L\"uscher,
Commun. Math. Phys. 85 (1982) 39

\bibitem{TopB}
A. V. Phillips and D. A. Stone,
Commun. Math. Phys. 103 (1986) 599;
{\it ibid}\/ 131 (1990) 255


\bibitem{Fujikawa}
K. Fujikawa,
A continuum limit of the chiral Jacobian in lattice gauge theory,
\hfill\break hep-th/9811235 

\bibitem{Suzuki}
H. Suzuki,
Simple evaluation of chiral jacobian with overlap Dirac operator,
\hfill\break hep-th/9812019 

\bibitem{Adams}
D. H. Adams,
Axial anomaly and topological charge in lattice gauge theory
with overlap Dirac operator,
hep-lat/9812003


\bibitem{KikukawaYamada}
Y. Kikukawa and A. Yamada,
Phys. Lett. B448 (1999) 265 

\bibitem{Chiu}
T.-W. Chiu, 
Phys. Lett. B445 (1999) 371

\bibitem{ChiuHsieh}
T.-W. Chiu and T.-H. Hsieh, 
Perturbation calculation of the axial anomaly of Ginsparg-Wilson fermion,
hep-lat/9901011


\bibitem{Kaplan}
D. B. Kaplan,
Phys. Lett. B288 (1992) 342;
Nucl. Phys. B (Proc. Suppl.) 30 (1993) 597

\bibitem{Shamir}
Y. Shamir, 
Nucl. Phys. B406 (1993) 90


\bibitem{AlvarezEtAl}
L. Alvarez-Gaum\'e, S. Della Pietra and V. Della Pietra,
Phys. Lett. B166 (1986) 177;
Commun. Math. Phys. 109 (1987) 691

\bibitem{NielsBohr}
L. Alvarez-Gaum\'e and S. Della Pietra,
The effective action for chiral fermions,
in: Recent developments in quantum field theory
(Niels Bohr Centennial Conference, Copenhagen, 1985),
eds. J. Ambj{\o}rn et al. (North-Holland, Amsterdam, 1985)

\bibitem{BallOsborn}
R. D. Ball and H. Osborn,
Phys. Lett. B165 (1985) 410;
Nucl. Phys. B263 (1986) 245

\bibitem{Ball}
R. D. Ball,
Phys. Lett. B171 (1986) 435;
Phys. Rept. 182 (1989) 1

\bibitem{KaplanSchmaltz}
D. B. Kaplan and M. Schmaltz,
Phys. Lett. B368 (1996) 44


\bibitem{SuzukiAction}
H. Suzuki,
Gauge invariant effective action in abelian chiral gauge theory
on the lattice,
hep-lat/9901012


\bibitem{Witten}
E. Witten,
Phys. Lett. B117 (1982) 324;
Nucl. Phys. B223 (1983) 422

\bibitem{ElitzurNair}
S. Elitzur and V. P. Nair,
Nucl. Phys. B243 (1984) 205

\bibitem{BaerCampos}
O. B\"ar and I. Campos,
work in progress


\bibitem{BrandtEtAl}
F. Brandt, N. Dragon and M. Kreuzer,
Phys. Lett. B231 (1989) 263;
Nucl. Phys. B332 (1990) 224; 
{\it ibid}\/ B332 (1990) 250

\bibitem{DuboisVioletteEtAl}
M. Dubois-Violette, M. Henneaux, M. Talon and C.-M. Viallet,
Phys. Lett. B267 (1991) 81;
{\it ibid}\/ B289 (1992) 361

\bibitem{Dragon}
N. Dragon,
BRS symmetry and cohomology,
Lectures given at Saalburg Summer School (1995),
hep-th/9602163


\bibitem{Gilkey}
P. B. Gilkey,
Invariance theory, the heat equation, and the Atiyah-Singer index theorem,
2nd ed. (CRC Press, Boca Raton, 1995)

\endbibliography

\bye